\documentstyle[psfig]{mn}

\def\ap3m{AP$^3$M}

\def\hkpc{h^{-1}\,{\rm kpc}}
\def\hMpc{h^{-1}\,{\rm Mpc}}

\def\i{\relax\ifmmode{\rm i}\else\char16\fi}

\def\lesssim{{_ <\atop{^\sim}}}

\def\ea{et~al.~}                            

\def\lesssim{\mathrel{\hbox{\rlap{\hbox{\lower4pt\hbox{$\sim$}}}\hbox{$<$}}}}
\def\gtrsim{\mathrel{\hbox{\rlap{\hbox{\lower4pt\hbox{$\sim$}}}\hbox{$>$}}}}

\newcommand{\ApJ}[3]    {\mbox{#3, ApJ~{#1},~#2}}
\newcommand{\ApJS}[3]   {\mbox{#3, ApJ~Suppl.~{#1},~#2}}
\newcommand{\ApJL}[3]   {\mbox{#3, ApJ~Lett.~{#1},~#2}}

\newcommand{\MNRAS}[3]  {\mbox{#3, MNRAS~{#1},~#2}}

\newcommand{\NewA}[3]   {\mbox{#3, NewA~{#1},~#2}}

\newcommand{\astroph}[1]{\mbox{{astro-ph/#1}}}

\begin{document}

   \title[Two-body relaxation in cosmological simulations]
	{Two-Body Relaxation in Cosmological Simulations}

   \author[Binney J.J. \& Knebe A.]
          {James Binney$^1$
           and
		   Alexander Knebe$^{1,2}$,
           \\
           $^1$Theoretical Physics, 1 Keble Road, Oxford OX1 3NP\\
           $^2$Centre for Astrophysics \& Supercomputing, Swinburne University, 
               Mail \# 31, PO Box 218, Hawthorn, VIC 3122, Australia
          }

   \date{Received ...; accepted ...}

   \maketitle

\begin{abstract}
It is logically possible that early two-body relaxation in simulations of
cosmological clustering influences the final structure of massive clusters.
Convergence studies in which mass and spatial resolution are simultaneously
increased, cannot eliminate this possibility.  We test the importance of
two-body relaxation in cosmological simulations with simulations in which
there are two species of particles. The cases of two mass ratios, $\surd2:1$ and
$4:1$, are investigated. Simulations are run with both a spatially fixed softening
length and adaptive softening using the publicly available codes GADGET and
MLAPM, respectively.

The effects of two-body relaxation are detected in both the
density profiles of halos and the mass function of halos. The effects
are more pronounced with a fixed softening length, but even in this
case they are not so large as to suggest that results obtained with
one mass species are significantly affected by two-body relaxation.

The simulations that use adaptive softening are  less affected
by two-body relaxation and produce slightly higher central
densities in the largest halos. They run about three times faster than
the simulations that use a fixed softening length.
\end{abstract}


\section{Introduction}
The Cold Dark Matter (CDM) model for the formation of cosmological
structure has enjoyed considerable success over the last decade and a
half. Recent results from the PSCz survey (Hamilton~\& Tegmark 2002)
underline the ability of this model to account successfully for
observations of the distribution of galaxies and the structure of rich
galaxy clusters within constraints set by measurements of the cosmic
background radiation, the theory of primordial nucleosynthesis and
observations of distant type Ia supernovae.

By contrast with these successes where large-scale phenomena are concerned,
there are serious doubts as to whether the CDM model is compatible with the
internal structures of galaxies (e.g.,  Spergel \& Steinhardt 2000). All these
difficulties can be traced to the cuspy central density profiles of dark
matter halos. This phenomenon was first identified by Navarro, Frenk \&
White (1996), who concluded that the central density diverged with radius as
$r^{-1}$. Subsequent work by Moore et al.\ (1998, 1999), Ghigna et al.\ (2000)  and
Klypin et al.\ (2001) suggests that the divergence is even steeper: $\rho\sim
r^{-1.5}$.

On sufficiently small scales, discreteness effects must cause deviations
between density profiles in simulations and those of real dark-matter halos,
in which the particle mass is believed to be tiny. Determining the smallest
scale on which numerical simulations are trustworthy has proved difficult
and has given rise to some controversy. Kravtsov et al.\ (1998), Ghigna et al.
(2000) and Klypin et al.\ (2001) have approached this problem by progressively
increasing the resolution of simulations started from initial conditions
that sampled the same underlying density field.  Kravtsov et al.\ (1998) argued that
their simulations were converging on a central density profile of halos that
was less cuspy than the NFW profile. Subsequently, Ghigna et al.\ (2000) and
Klypin et al.\ (2001) showed that simulations  converge on profiles more
cuspy than the NFW profile if one simultaneously increases both the mass and
the spatial resolution; Kravtsov et al.\ (1998) had increased only the spatial
resolution. 

While the results of Ghigna et al.\ (2000) and Klypin et al.\ (2001) suggest that real dark
matter halos should have very cuspy centres as Moore et al.\ (1998) originally
concluded, they do not establish this claim beyond all reasonable doubt. The
reason is that when structure forms bottom-up, as in the CDM paradigm, and
mass and spatial resolution are increased together, the first virialized
systems are few-body systems, regardless of the resolution employed. In such
systems the two-body time is comparable to the dynamical time, and two-body
interactions tend to make the core denser, and less vulnerable to subsequent
tidal destruction. As clustering progresses, one can imagine that the
dynamics is dominated by the interaction of quasi-particles formed by
first-generation systems. The early clustering of these quasi-particles is
again heavily influenced by two-particle interactions, and leads to the
formation of a new generation of quasi-particles, each of which is made up
of quasi-particles of the previous generation. Thus one can imagine the
dynamics right up to the macroscopic scale being influenced by the two-body
interactions that took place between the real simulation particles as the
density field first went non-linear. Convergence tests of the type performed
by Ghigna et al.\ (2000) and Klypin et al.\ (2001) would be powerless to expose discreteness
effects if this picture were valid, because a low-resolution simulation
would simply enter the evolutionary sequence of a high-resolution simulation
at the stage proper to the mass of its simulation particles.

An indication that we should worry about the possibility just described
comes from the observation that cuspy profiles are only obtained when the
smoothing kernel used to calculate gravitational forces from particle
positions is hard. Now, Fig.~10 of Knebe, Green \& Binney (2001) shows that
for the values of the softening parameter that yield cuspy profiles, the net
gravitational force on a particle may be dominated by the force from its
nearest neighbour prior to the virialization of the smallest modelled
scales. This circumstance is worrying, because the essence of collisionless
dynamics is that the forces acting on each particle reflect the mean density
field rather than the chance location of a neighbour.  Moreover, large
forces from neighbours can affect the dynamics of particles more strongly at
early times, when random velocities are small and thus force contributions
are relatively slowly changing.

The standard way of determining the significance  for
a simulation of two-body relaxation is to include particles of more than one mass: if the
simulation is collisionless, the final distributions of the particles will
be independent of mass, whereas the more massive particles will tend to sink
to the bottoms of potential wells if two-body relaxation is significant.
Since no results from multi-mass cosmological simulations have appeared, we
present here the results of such experiments. The experiments were conducted
with two radically different simulation codes, the tree-code GADGET
(Springel, Yoshida~\& White 2001) and the adaptive multigrid code, MLAPM
(Knebe et al.\ 2001). We detect the effects of two-body relaxation in
results produced with both codes. However, the magnitude of the effect is
not so large as to suggest that the cuspiness of dark-matter halos is
generated by two-body relaxation. We argue instead that cuspiness arises
from violent relaxation, and will invariably arise when gravity causes
particles to cluster collisionlessly in a cosmological context.

\section{Initial conditions}

 Cosmological simulations are invariably initialized by using the Zel'dovich
approximation to displace particles from an initial approximately
homogeneous distribution (e.g., Efstathiou~\ea 1985). The simplest way to
approximate a homogeneous distribution is to place particles at the vertices
of a regular grid. Since the regularity of the grid obviously induces
unwanted long-range correlations between the particles, another
computationally more demanding way of producing an approximately homogeneous
mass distribution has been widely used. This technique
involves scattering particles at random over the computational volume, and
then moving them for some time under the influence of a fictitious repulsive
gravity. The motion of particles away from one another flattens out
Poissonian density inhomogeneities of the initial distribution to produce a
rather homogeneous glass (White, 1996).

\begin{figure}
\centerline{\psfig{file=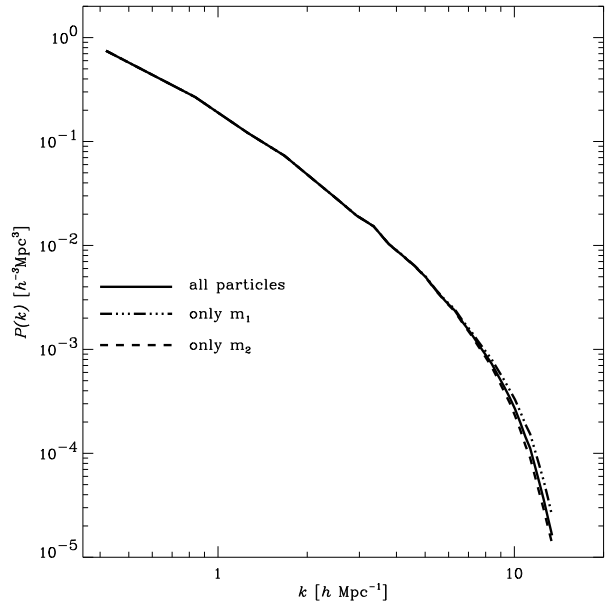,width=\hsize}}
\caption{Power spectra of the initial particle distributions. \label{initP}}
\end{figure}

\begin{figure*}
\centerline{\psfig{file=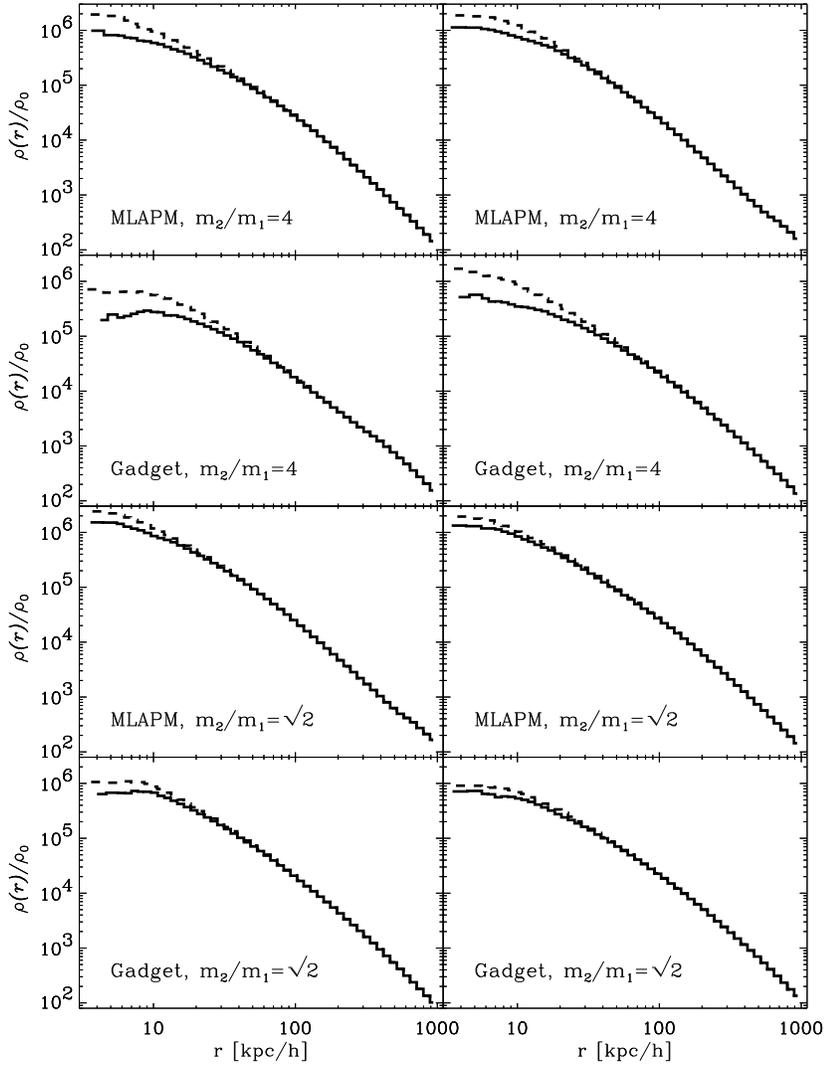,width=.6\hsize}}
\caption{Density profiles of the two most massive halos in four simulations.
Each column shows results for one of the halos. Full curves show the
number density of the less massive particles, and dashed curves show the
number density of the more massive particles.\label{massprofs}}
\end{figure*}

We have initialized simulations containing two particle species using both
of the above approaches to the production of an approximately homogeneous
distribution. Unfortunately, glasses containing two particle species contain
complex short-range correlations because the size of the void around each
particle that is generated by repulsive gravity depends on the mass of the
particle. Consequently, multi-mass glasses have complex and mass-dependent
short-range correlations that obscure the effects of two-body
relaxation. Since initial conditions based on a regular lattice yield much
cleaner results, we confine the discussion to these simulations. The results
obtained with glass-initialized simulations are consistent with the results
presented here.

Our lattice based simulations were initialized as follows. 
$64^3$ particles of one species were placed on the nodes of a 
regular grid with $64^3$ cells covering a physical scale of 
$15 \hMpc$ on a side.
Then $64^3$ particles were placed on the nodes of the grid that is offset from
the first grid by half a lattice spacing parallel to each axis. Finally
Zel'dovich displacements were applied to each mass. In the final
configuration, the particles of a given species define a density field that
differs only in its overall normalization from the density field defined by
particles of the other species. Hence the composite density field defined by
both species together is also the same, up to an overall normalization. The
masses of individual particles were set such that the composite density was
that of the standard CDM model and the masses of the two species were in the
ratio $1:\surd2$ or $1:4$. It follows that the more massive species
accounted for 59 per cent and 80 per cent of the total mass in the two
cases. Fig.~\ref{initP} shows the power spectra of the initial particle
distribution. The power spectra of the light ($m_1$) and the heavy particles
are indistinguishable except at the highest wavenumbers, where the offset
between the lattices from which the particles started gives rise to very
slight differences in the way in which artifacts arising from the lattices
impact on the power spectrum.

\section{Results}

We evolved the particle distributions that are described by Fig.~\ref{initP}
in a standard CDM cosmology from redshift $z=35$ to the current epoch using
both the tree code GADGET (Springel et al.\ 2001) and our own multigrid
code, MLAPM (Knebe et al.\ 2001). The essential difference between these
codes is that whereas GADGET uses a spatially invariant gravitational
softening length, MLAPM softens gravity adaptively so as to provide a
compromise between enhanced force resolution and reduced particle noise in
the force-field. The GADGET's Plummer softening length was set to $\epsilon
= \min[5,1/a(t)]\hkpc$, where $a<1$ is the scale factor, so at late times
the code used accurately Newtonian forces for distances greater than
$5h^{-1}\,$kpc.
MLAPM operates by refining the grid on which it
solves Poisson's equation whenever the density exceeds a critical
value $\rho_{\rm ref}$, typically set to 8 particles per cell.  For
these experiments the density compared to $\rho_{\rm ref}$ was not the
density of gravitating mass, but the number density of particles,
irrespective of their mass. The finest grid employed at the end of the
simulation had a mesh spacing equivalent to $8192^3$ cells in the
computational volume, each cell being $1.8\hkpc$ on a side. On such a mesh the force law becomes Newtonian for
distances $\ga5\hkpc$ (Knebe et al., 2001). Hence the spatial resolution
provided by the two codes agreed well in high-density regions.

The GADGET runs required $244\,398$ and $249\,677$ timesteps, while the
MLAPM runs used the equivalent of 1000 timesteps on the finest domain grid
(128 cells on a side), during which $64\,000$ timesteps would be taken on an
$8192^3$ grid.

Fig.~\ref{massprofs} shows the number-density profiles of the less and more
massive particles of the two most massive halos to form in the simulations.
As we will see below in Fig.~\ref{massrats}, both these halos
contain about an equal number of light and heavy particles in every run.
For each halo four pairs of profiles are shown, corresponding to the two
mass ratios and two codes. Since the full curves, which show the profile of
the less massive particles, always lie below the dashed curves, mass
segregation is clearly detected. The effect is slightly more pronounced for
the halo shown on the left than for that shown on the right. The main
difference between the results obtained with the two codes is that MLAPM
produces a central density that is about a factor 2 larger than that
produced by GADGET, and mass segregation is more pronounced in
the profiles produced by GADGET.

\begin{figure}
\centerline{\psfig{file=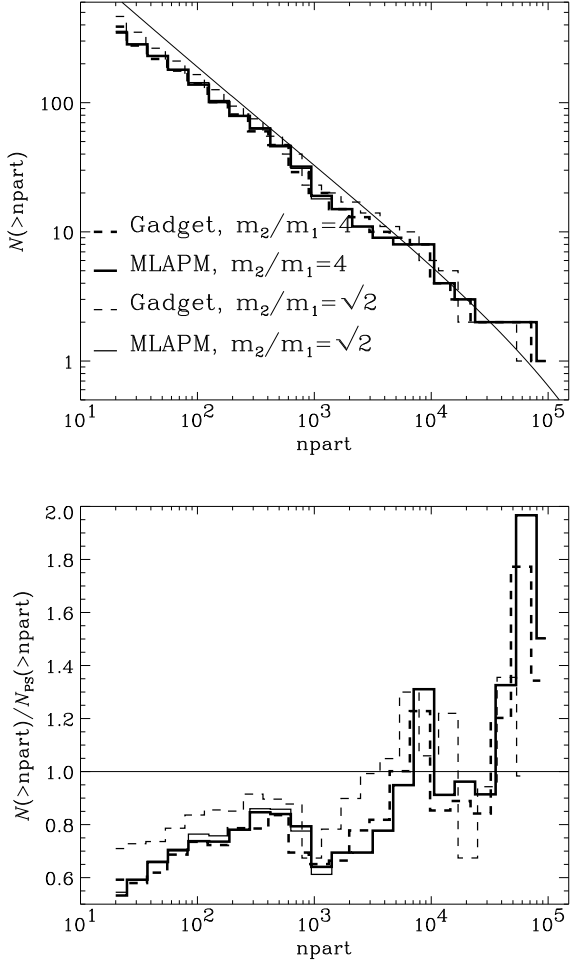,width=\hsize}}
\caption{Mass functions from the simulations. The friends-of-friends
algorithm has been used. The ordinate shows the number of particles in each
object, not the total mass. The lower panel shows the ratio of the actual
number of clusters to predicted by Press-Schechter theory, and the thin full
curve in the upper panel shows this prediction.\label{massfuncs}}
\end{figure}

The effects of two-body relaxation on the less massive halos in the
simulations is examined by Fig.~\ref{massfuncs}, which shows for the four
simulations the numbers of halos formed that contain in excess of a given
number of particles (irrespective of particle mass). The thin full curve in
the upper panel shows the prediction of Press-Schechter theory. In the lower
panel we show for each species the ratio of the actual number of cluster to
that predicted by Press-Schechter theory. The full histograms, which show
the results obtained for the two mass ratios by MLAPM, are almost
indistinguishable. The results obtained with GADGET agree moderately well
with the MLAPM results only for $m_2/m_1=4$. The GADGET results for
$m_2/m_1=\surd2$ show significantly more low-mass clusters.

\begin{figure}
\centerline{\psfig{file=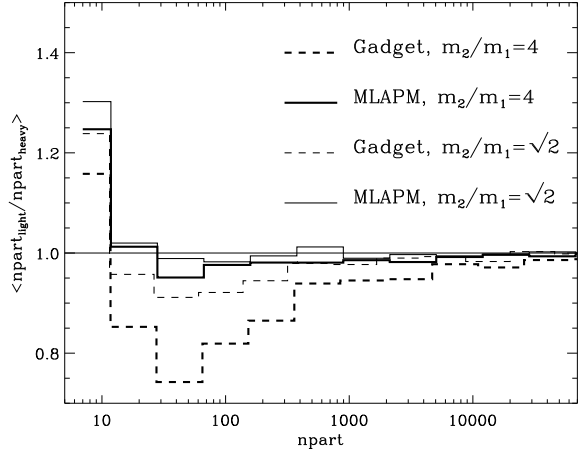,width=\hsize}}
\caption{Ratio of the number of light to heavy particles in halos as a
function of the total number of particles in the halo.\label{massrats}}
\end{figure}

Fig.~\ref{massrats} clarifies what is going on by plotting as a function of
the total particle number of a given halo the ratio of the number of light
to heavy particles.  Since in the simulation as a whole there are equal
numbers of particles of each species, this ratio would ideally be unity,
independent of total particle number. The results from MLAPM deviate
significantly from this ideal only for $n_{\rm part}<10$, where for both
mass ratios clusters have more light than heavy particles. The results from
GADGET, by contrast, reveal a significant tendency for clusters containing
$>10$ particles to have more heavy than light particles. The tendency is
strongest for $m_2/m_1=4$, as would be expected if it were the result of
two-particle relaxation causing heavy particles to sink towards the centres
of clusters, while light particles evaporate from them. It is not clear why
in all simulations clusters with only a handful of particles tend to have an
excess of light particles, but this phenomenon may be connected with the
fact that in Fig.~\ref{initP} the power spectrum of the light particles is
higher than that of the heavy particles at large $k$. We believe this
phenomenon to be of no physical significance and to arise by chance from the
phases of the waves used to displace the particles.

\section{Discussion}

Simulations of cosmological clustering with two mass species have been used
to explore the importance of two-body relaxation in general cosmological
simulations.  In the two largest halos that form in our simulations, mass
segregation is a small but detectable effect in the sense that the
difference in the central densities of particles with masses in the ratio
$4:1$ is comparable to the difference in the central densities obtained for
any one species with the two codes (about a factor 2). Mass segregation is
stronger in simulations run with GADGET, which has (spatially) fixed
softening, than with MLAPM, which softens adaptively.

In the MLAPM simulations there is no evidence that two-particle relaxation
enhances the fraction of heavy particles in clusters. The GADGET simulations
show a clear tendency for clusters with more than a handful of particles to
have more massive than light particles. This tendency increases in strength
with the mass ratio $m_2/m_1$, just as is expected if it is driven by
two-particle relaxation.

Given the clear indication that two-particle relaxation is more pronounced
in the GADGET simulations, it is interesting that it is in the MLAPM
simulations that the central densities of the two most massive clusters are
largest (by about a factor 2).

The major difference between the GADGET and MLAPM simulations was one
of computational cost: using a small softening length everywhere
increases the time required to run a simulation, both because
it increases the cost of evaluating forces, and, more importantly,
because it forces one to smaller timesteps. Quantitatively, the
present GADGET simulations took longer than the corresponding MLAPM
simulations by a factor about three. Since the quality of the
scientific output from an N-body simulation is always limited by the
number of particles employed, and most simulators are more limited by
the availability of computer time than computer memory, the different
speeds of our simulation amounts to a strong case for the use of
adaptive softening of the type that MLAPM, and certain tree codes
(Dehnen 2000) provide.

The equipartition time, on which mass segregation develops, is faster than
the two-body relaxation time by roughly the ratio of the particle masses.
By contrast, the core-collapse timescale, on which the central density of a
halo is increased by two-body relaxation, is typically tens to hundreds of
two-body relaxation times (e.g., Binney \& Tremaine 1987).  Hence, our
finding that mass segregation is a marginal effect for mass ratio $4:1$,
strongly suggests that the cuspiness of dark-matter halos in simulations is
not an artifact produced by two-body relaxation. This conclusion is
reinforced by the observation that central densities are highest in
simulations in which mass segregation is least evident.

If the cuspiness of halos is not caused by high-$k$ power in the input
spectrum, as experiments with WDM suggests (Knebe et al.\ 2002), and not
produced by two-body relaxation as we have argued, what {\it is\/} its
origin? It must result from violent relaxation: in the absence of high-$k$
power, the quasi-linear regime will continue until massive objects are ready
to virialize. These will collapse by large factors because they will
collapse from smooth and symmetrical initial conditions. Numerical
experiments long ago established that the peak density in the final
virialized entity is comparable to the peak density achieved during collapse
(Doroshkevich \& Klypin 1981; van Albada 1982). Hence, there is an argument
for less high-$k$ power giving rise to higher central densities.

In the presence of high-$k$ power, violent relaxation will be unimportant
for the formation of massive objects: these will form hierarchically through
a series of mergers, and their cores are likely to be dominated by material
that virialized early on as part of a low-mass object. The insensitivity of
density profiles to the initial power spectrum must be the result of a
conspiracy, which ensures that as high-$k$ power is smoothed away, the loss
of high-density seeds from early virialization is counteracted by an
increase in the effectiveness of violent relaxation at late times. The
bottom line is that cuspy density profiles must be considered inevitable so
long as structure formation is dominated by a collisionless fluid.

\section*{Acknowledgments}
We benefited from a valuable conversation with A. Klypin.



\begin{thebibliography}{}

\bibitem[binney0]{binney0} 
        {Binney, J., \& Tremaine, S. 1987, {\em Galactic Dynamics} 
         (Princeton: Princeton University Press)}







\bibitem[dehnen1]{dehnen1}
        {Dehnen W., \ApJL{536}{39}{2000}}

\bibitem[dorosh]{dorosh}
		{Doroshkevich, A.G., Klypin, A.A., 1981, Soviet Astr., 25, 127}

\bibitem[Efstathiou1]{efstathiou1}
        {Efstathiou G., Davis M., Frenck C.S., White S.D.M., \ApJS{57}{241}{1985}}

\bibitem[Ghigna]{ghigna1}
        {Ghigna S., Moore B., Governato F., Lake G., Quinn T., Stadel J.,
         \ApJ{544}{616}{2000}}

\bibitem[hamilton1]{hamiton1}
        {Hamilton A., Tegmark M., 2002, MNRAS in press \astroph{0008392}}

\bibitem[klypin99]{klypin99}
        {Klypin A.A., Kravtsov A.V., Valenzuela O., Prada F.,
         \ApJ{522}{82}{1999}}

\bibitem[klypin01]{klypin2}
        {Klypin A., Kravtsov A.V., Bullock J.S., Primack J.R.,
         \ApJ{554}{903}{2001}}

\bibitem[knebe1]{knebe1}
        {Knebe A., Devriendt J.E.G., Mahmood A., Silk J.,  2002, MNRAS, 329
	, 813}

\bibitem[knebe2]{knebe2}
        {Knebe A., Green A., Binney J.J.,  \MNRAS{325}{845}{2001}}

\bibitem[kravtsov]{kravtsov}
        {Kravtsov A.V., Klypin A., Bullock J.S., Primack J.R.,
         \ApJ{502}{48}{1998}}

\bibitem[moore0]{moore0}
        {Moore B., Governato F., Quinn T., Stadel J.,
         Lake G., \ApJ{499}{L5}{1998}}

\bibitem[moore1]{moore1}
        {Moore B., Quinn T., Governato F., Lake G.,  Stadel J.,
         Lake G., \MNRAS{310}{1147}{1999}}

\bibitem[NFW]{NFW}
		Navarro J.F., Frenk C.S., White S.D.M., 1996, ApJ, 462, 563


\bibitem[spergel1]{spergel1}
	{Spergel D.N., Steinhardt P.J.,  2000, Phys. Rev. Lett. {\bf 84}, 3760}

\bibitem[Springel1]{springel1}
        {Springel V., Yoshida N., White S.D.M., \NewA{6}{79}{2001}}

\bibitem[Tjiert]{Tjiert}
		{van Albada, T.S., \MNRAS{201}{939}{1982}}



\bibitem[white1]{white1}
        {White S.D.M., 1996, {\em Cosmology and Large-Scale Structure,
         Les Houches Session LX}, eds. Schaeffer R., 
         Silk J., Spiro M., Zinn-Justin J., Elsevier 1996, p. 349}

\end{thebibliography}
\end{document}